\documentclass[12pt]{article}

\usepackage{scicite}
\usepackage{graphicx}
\usepackage{times}
\usepackage{caption}
\usepackage{amsmath}
\usepackage{braket}

\newenvironment{sciabstract}{%
\begin{quote} \bf}
{\end{quote}}

\title{Microscopy with undetected photons in the mid-infrared}  

\author
{  \parbox{\linewidth}{\centering Inna Kviatkovsky$^{1\ast}$, Helen M. Chrzanowski$^{1}$, Ellen G. Avery$^{2,3,4,5,6,7}$, Hendrik Bartolomaeus$^{2,3,4,5,6}$, Sven Ramelow$^{1,8}$ }\\
\\
 \parbox{\linewidth}{\centering \normalsize{$^1$ Institut f\"{u}r Physik, Humboldt-Universit\"{a}t zu Berlin, Berlin, Germany.}\\
\normalsize{$^2$ Experimental and Clinical Research Center, a cooperation of Charité - Universitätsmedizin Berlin and Max Delbrück Center for Molecular Medicine, Berlin, Germany.}\\
\normalsize{$^3$ Charité-Universitätsmedizin Berlin, corporate member of Freie Universität Berlin, Humboldt-Universität zu Berlin, and Berlin Institute of Health, Berlin, Germany.}\\
\normalsize{$^4$ Max Delbrück Center for Molecular Medicine in the Helmholtz Association, Berlin, Germany.}}\\
\normalsize{$^5$ DZHK (German Centre for Cardiovascular Research), partner site Berlin, Germany.}\\
\normalsize{$^6$ Berlin Institute of Health (BIH), Berlin, Germany.}\\
\normalsize{$^7$ Freie Universität Berlin, Germany.}\\
\normalsize{$^8$ IRIS Adlershof, Humboldt-Universität zu Berlin, Berlin, Germany.}\\
\\
\normalsize{$^\ast$Corresponding author; E-mail:  innakv@physik.hu-berlin.de.}
}

\date{}

\begin{document}

\baselineskip24pt

\maketitle 

\begin{sciabstract}
  Owing to its capacity for unique (bio)-chemical specificity, microscopy with mid-IR illumination holds tremendous promise for a wide range of biomedical and industrial applications. The primary limitation, however, remains detection; with current mid-IR detection technology often marrying inferior technical capabilities with prohibitive costs. This has lead to approaches that shift detection to wavelengths into the visible regime, where vastly superior silicon-based camera technology is available. Here, we experimentally show how nonlinear interferometry with entangled light can provide a powerful tool for mid-IR microscopy, while only requiring near-infrared detection with a standard CMOS camera. In this proof-of-principle implementation, we demonstrate intensity imaging over a broad wavelength range covering  3.4-4.3 $\mu$m and demonstrate a spatial resolution of 35 $\mu$m for images containing 650 resolved elements. Moreover, we demonstrate our technique is fit for purpose, acquiring microscopic images of biological tissue samples in the mid-IR. These results open a new perspective for potential relevance of quantum imaging techniques in the life sciences.
\end{sciabstract}

\section*{Introduction}

Mid-IR imaging and microscopy is extensively used in various fields such as biology and medicine\cite{Kwak2011,Evans,Evans2007,Bellisola2012,Potter2001,Miller2013}, environmental sciences\cite{Tagg2015} and microfluidics\cite{Chrimes2013}. 
Sensing with mid-IR light exploits the distinct rotational and vibrational modes of specific molecules\cite{Shaw1999}. This spectral fingerprint can be used as a contrast mechanism for mid-IR imaging, circumventing the need for labelling. Such non-invasive and label-free imaging techniques are especially important for bio-imaging procedures, as they permit the observation of largely unaltered living tissues. The current state-of-the-art mid-IR imaging technique is Fourier transform infrared spectroscopic (FTIR) imaging\cite{Bhargava2012}. It heavily relies on infrared (IR) technologies, namely broadband IR sources and detectors. While the gap in technology and price between IR and visible sources is slowly closing\cite{Godard2007}, IR detection technology lags significantly behind its visible counterparts\cite{Rogalski2002,Scribner1991,Rogalski2016}, such as complementary metal-oxide-semiconductor (CMOS) and charge-coupled device (CCD) technologies. Furthermore, IR detectors are costly and technically challenging, often requiring cryogenic cooling, and moreover are subject to severe export restrictions due to dual-use issues. 

In order to bypass the need for IR detectors, techniques such as coherent Raman anti-stokes scattering (CARS) microscopy\cite{Mu2002,Cheng2015} were developed. Here, the weakness of the Raman-effect and intrinsic noise mechanism require high laser intensities and only permit slow point-by-point scanning. Photothermal lens microscopy\cite{zhang2016depth} and photoacoustic microscopy\cite{shi2019high} in the mid-IR are two imaging modalities that are capable of imaging fresh biosamples with high spatial resolution. Nevertheless, raster scanning is required and sample illumination is comparatively invasive. Other approaches employ frequency conversion to shift the detection frequency to the visible, while still sensing in the IR region for the highly desired information. Up-conversion methods have demonstrated imaging in the near and mid-IR\cite{Demur2018,Dam2012,Junaid2019}, but conversion efficiency and the number of converted spatial modes remain a significant challenge, especially as they scale unfavourably with each other.

\begin{figure}[h!]
\centering
\includegraphics[width=14cm]{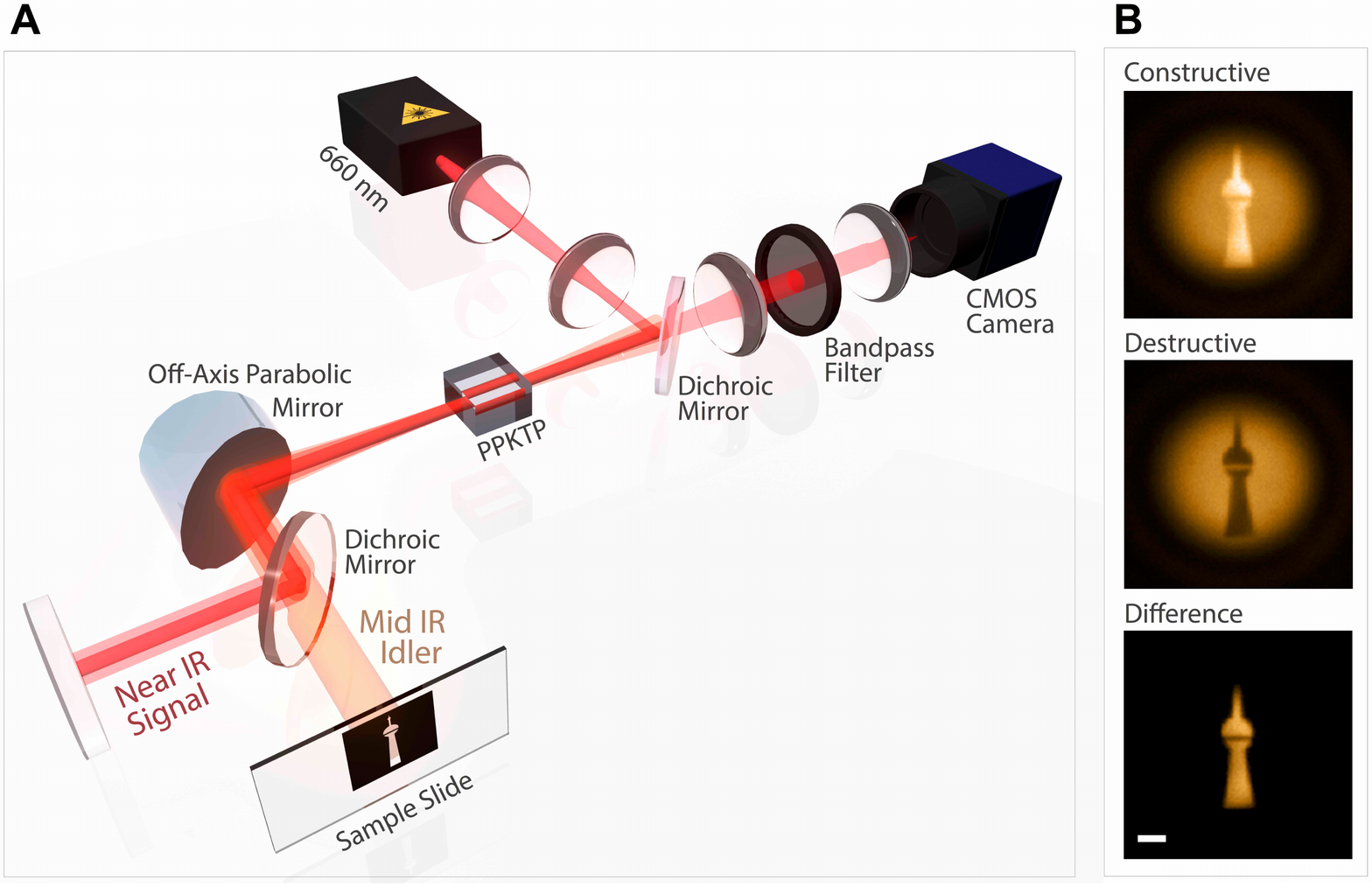}
\caption*{{\bf Fig. 1. (A) Experimental Setup:} A 660 nm CW laser pumps a highly non-degenerate SPDC process. The signal and idler fields generated on the first pass of the 2 mm ppKTP crystal are split via a dichroic mirror. The sample to be imaged is placed in the Fourier plane of the idler, which coincides with its end mirror. Both the idler and signal fields are reflected back, recombined and back propagated into the nonlinear crystal with the coherent pump field. The resulting signal field is imaged on a CMOS camera. {\bf (B)} Constructive, destructive and difference interference images of the signal for a cardboard cutout probed by the mid-IR idler. The scale bar corresponds to 2 mm.}
\end{figure}

A drastically different approach utilizes the interference of an entangled photon-pair with widely different wavelengths, and requires neither laser sources nor detectors at the imaging wavelength. The initial proof-of-concept for wide field imaging was demonstrated\cite{Lemos2014} at 1550 nm with moderate spatial resolution. There the approach was based on induced coherence without induced emission\cite{Wang,Zou1991}, but similar effects can be realized in general by nonlinear interferometers\cite{Chekhova2016}. It has also seen use, albeit in a single spatial mode regime for other modalities of sensing, such as spectroscopy\cite{Kalashnikov2016}, refractometry\cite{Paterova:2018dz} and optical coherence tomography\cite{Valles:2018df, Paterova:2018kla}. 

In this work, we show for the first time how highly multi-mode quantum nonlinear interferometry forms a powerful tool for microscopic imaging in the mid-IR using only a medium powered visible laser and a standard CMOS camera. We also derive explicit formulas for the field-of-view (FoV) and resolution for wide-field imaging with highly non-degenerate photon pairs, which are verified experimentally and numerically reproduced using a full quantum formalism \footnote{See supplementary information}.

\begin{figure}[h!]
\centering
\includegraphics[width=16cm]{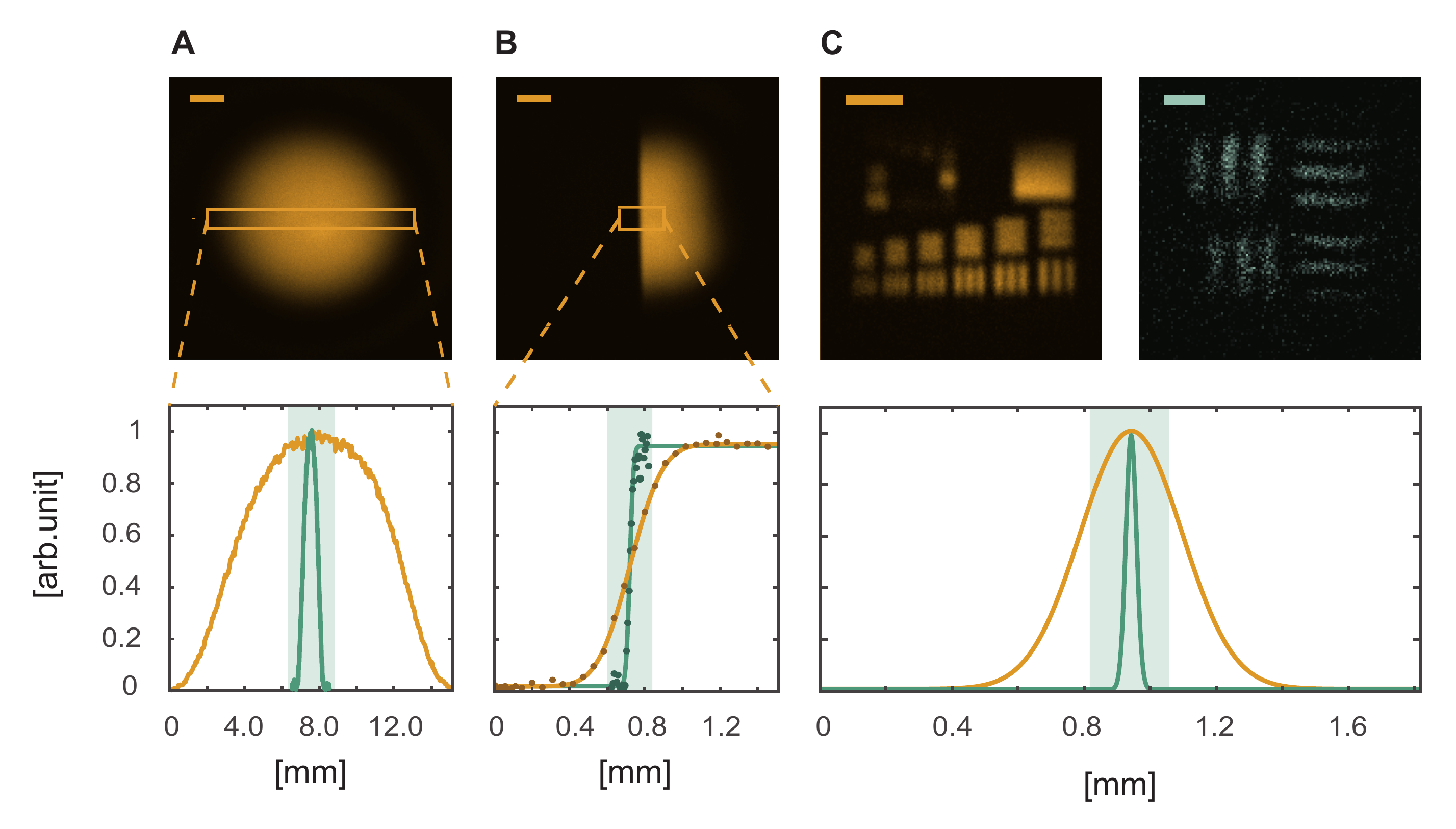}
\caption*{{\bf Fig. 2. Characterization of the imaging arrangements.} The images and data of the unmagnified and magnified setups are presented in orange and green respectively. ({\bf A}) Measured FoV of the unmagnified and magnified setups is $9100 \pm 82\mu$m and $819 \pm 9\mu$m respectively. ({\bf B}) The edge response functions fitted to the data of the two imaging arrangements. ({\bf C}) Measured resolution of the unmagnified and magnified setups is $322 \pm 5 \mu$m and $35 \pm 5 \mu$m respectively. The smallest features in a resolution target that can be resolved for each arrangement are presented. Orange scale bar corresponds to 2 mm, green scale bar corresponds to 0.1 mm. Unmagnified (magnified) images were acquired with 1 second integration time and 200 (400) mW pump power.}
\end{figure}

In our implementation, a nonlinear interferometer is formed by double-passing a periodically poled potassium titanyl phosphate (ppKTP) crystal in a folded Michelson geometry\cite{Chekhova2016}. As the pump passes the crystal twice, it can generate a single pair of signal and idler photons via spontaneous parametric down conversion (SPDC) in the first and/or second crystal. The signal and idler modes after the first pass of the crystal are aligned such, that when propagating back for the second pass they perfectly overlap with the signal and idler modes for the second possibility for the bi-photon generation. This results in indistinguishability, and thus interference of single bi-photons generated in the `first' and/or `second' crystal. The interference (of only one bi-photon with itself) can be fully measured by solely looking at the signal photons with a CMOS-camera. It reveals the phase and absorption an idler photon would experience after the first pass. Remarkably, no complex or cost-intensive components are required to realize such a setup.

In this work the nonlinear crystal was engineered for highly non-degenerate signal and idler wavelengths. Using broadband phase matching\cite{Vanselow2019Ultra-broadbandPairs} the idler wavelength can be selected in a large range between 3.4-4.3 $\mu$m at room temperature, while the corresponding signal wavelength is in the 780-820 nm range.  The strong spatial correlations between the signal and idler modes ensure that any distinguishing information obtained by the idler field between the first and second pass of the crystal, will be encoded onto the interference of the near-IR light after the second crystal. This allows the simultaneous retrieval of both, spatially resolved phase and amplitude information of a sample put into the idler arm. We characterise the mid-IR imaging properties of this system with an off-the-shelf CMOS camera. Moreover, the ability to manipulate the FoV and accordingly, the system resolution is demonstrated. Specifically, using a x10 magnification, details down to 35 microns are shown to be detected, which we use for acquiring microscopic images of a biological sample. 

\section*{Results}

The experimental setup is detailed in Fig. 1. The initial characterisation of the imaging technique was made in an unmagnified configuration, with both the end mirrors of the interferometer placed at the far-field of the crystal. The sample to be imaged is placed on the idler mirror. Whilst the unmagnified configuration possesses limited spatial resolution, it provides a straightforward means to characterise the imaging capacity of the system. The lateral resolution was ascertained by measuring the spatial response to an edge knife (Fig. 2B), yielding $322\pm 5 \mu$m. The estimated number of spatial modes is $800 \pm 20 $. In addition, a USAF clear path resolution target was illuminated (Fig. 2C). These values are consistent with the theoretical model \footnote{See supplementary information} developed using a theoretical framework generalized from that of ghost imaging\cite{Moreau2018}. 

\hspace{10pt}
\begin{figure}[h!]
\centering
\includegraphics[width=14cm]{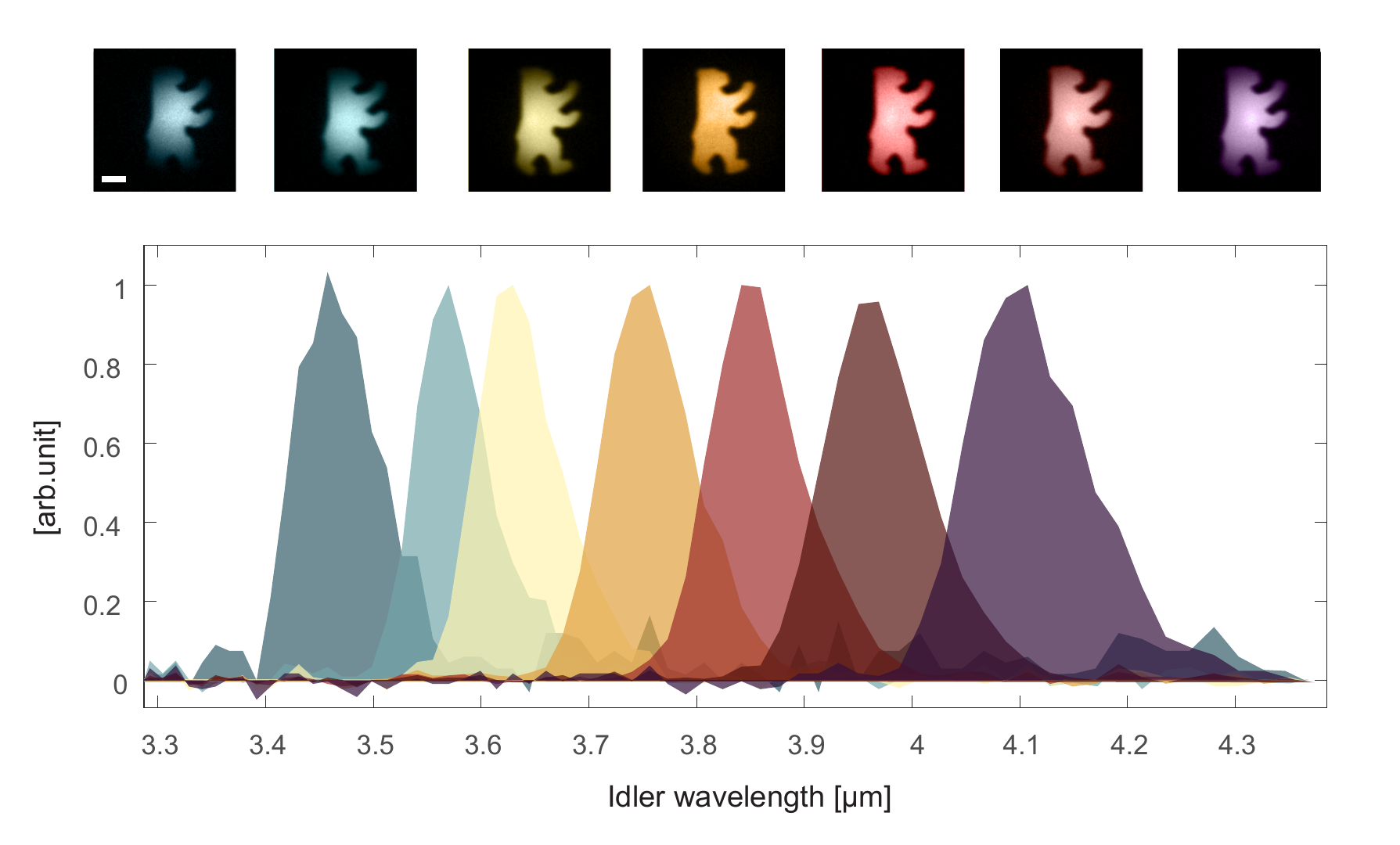}
\caption*{ {\bf Fig. 3. Multi-spectral imaging:} Obtained signal transmission images for varying mid-IR illumination wavelengths. The scale bar corresponds to 2 mm. The spectra were recorded at the signal wavelength with a grating spectrometer and converted to the corresponding mid-IR wavelength.}
\end{figure}

To demonstrate microscopy, a 10 fold magnification was realised via a telescope in the idler arm. Characterisation of the magnified setup was performed in a similar manner, with the results summarised in Fig. 2. The obtained number of spatial modes is $655 \pm 57 $. The system resolution of $35\pm 5 \mu$m is below the smallest available line-pair for our clear path resolution target (Fig. 2C). 

The number of measured spatial modes in the unmagnified (magnified) realisation are about 88\% (72\%) of the theoretical value. We attribute this reduction in both optical arrangements to alignment imperfections, namely, in matching the corresponding imaging planes precisely, as well as chromatic aberrations. In addition, the smaller depth of focus in the magnified regime brings higher sensitivity to mismatch of imaging and sample planes.

Our built-for-purpose SPDC source yields approximately $10^8$ pairs/s within the filter bandwidth per 400 mW of pump light, which are distributed over the around 18.000 pixels covering the FoV on the CMOS camera. This results in approximately 5000  photons/pixel/s, far above the intrinsic camera noise, and leads to the shot-noise and visibility limited signal-to-noise ratios (SNR) of the unmagnified (magnified) realisation of around 25 (8) that we observe in our measurements. This corresponds to a resolvable transmission difference of 4\% (12\%). The difference between the two realisations stems from the reduced visibility of the magnified arrangement, which we attribute to technical imperfections such as additional losses, alignment imperfections and aberrations.
\hspace{10pt}
\begin{figure}
\centering
\includegraphics[width=16cm]{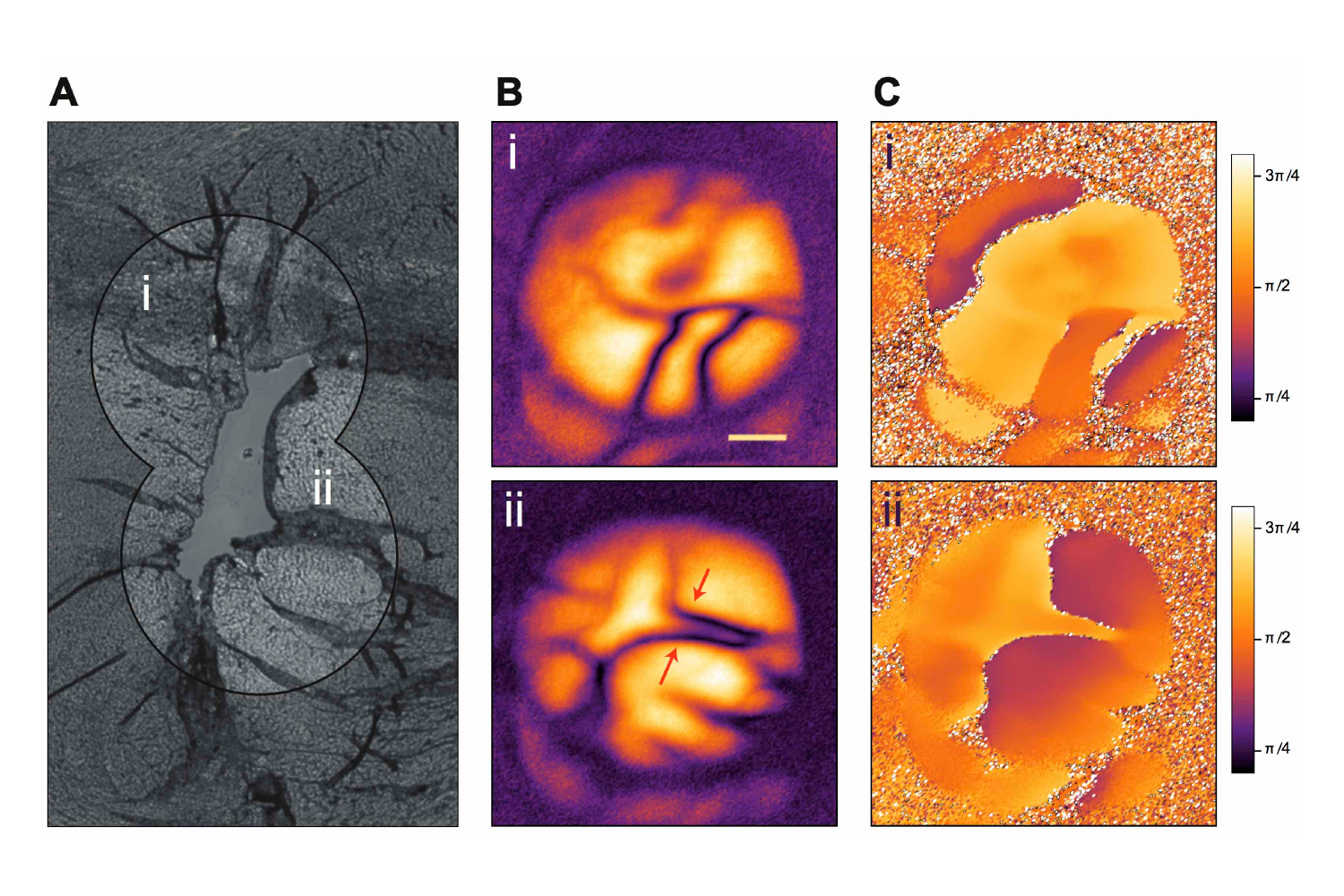}
\caption*{{\bf Fig. 4. Bioimaging} Imaging of a histology sample of a mouse heart with ({\bf A}) bright field microscopy, and absorption ({\bf B}) and phase imaging ({\bf C}) with mid-IR quantum light. Scale bar corresponds to $200 \mu$m. Images were reconstructed by averaging 10 images at 1 s integration time for 15 axial positions within the coherence length of the bi-photon. Pump power was 400 mW corresponding to a sample illumination power of less than 20 pW.}

\end{figure}
The highly broadband nature of the down-conversion source combined with the tight energy correlations shared between signal and idler on the order of 1 MHz (given by the pump laser linewidth) easily facilitates hyper-spectral imaging. In our proof-of-concept demonstration we use a tuneable interference filter with 3.5nm bandwidth immediately prior to detection. Fig. 3 presents transmission images obtained with the sample illumination spanning from 3.4-4.3 $\mu$m with corresponding spectral selection of around 85 nm in the mid-IR. Note that the spectral resolution can be straightforwardly enhanced by narrower filtering and trading-off signal strength. 

Finally, the potential of the presented method for realistic, biological samples was demonstrated using an unstained histology sample of a mouse heart (see Methods for preparation) (Fig. 4) The mid-IR images (Fig. 4B) were obtained by axially scanning the interferometer displacement inside the coherence length and extracting the visibility and phase of the interference signal for each pixel. This eliminates any ambiguity between loss and destructive interference (owing to a non-uniform phase front) that may arise in a single shot measurement. It also permits straightforward reconstruction of the wide-field phase-contrast images (Fig. 4C). 
In the resulting images of the mouse heart in Fig. 4 the arrows indicate a portion of the endocardium, the innermost layer that lines the heart ventricles. This can be seen as a dark purple region indicating high absorption. This layer separates the ventricle itself and the myocardium, the cardiac muscle that constitutes the bulk portion of the heart tissue. The images are a testament to the high tolerance of the presented imaging method to loss and scattering. 

\section*{Discussion}

We have demonstrated the capacity of mid-IR imaging via nonlinear interferometry for real world imaging tasks requiring only cost-efficient components. Via careful addition of a magnification step, we demonstrate imaging of feature sizes down to 35 $\mu$m. 

Further, owing to the use of a broadband SPDC, the extension to hyper-spectral imaging is uncomplicated. To demonstrate the real world promise of this method for non-destructive biological sensing, we have imaged a wet biological sample with a low sample illumination of less than 20 picowatts. 
Given this, one must contemplate what underlies the (perhaps surprising) efficacy of the presented technique. Firstly, due to the SPDC process, for every detected signal photon exists a corresponding idler photon, and, owing to the intrinsic unity efficiency of the nonlinear-interferometer, any image information carried by that idler photon can be transferred perfectly to the signal photon. Thus, in the absence of loss and mode mismatch, one can effectively consider the `mid-IR information detection efficiency' to be determined by the efficiency and noise properties of the near-IR camera used to image the signal field, with modern silicon technology permitting near shot-noise limited images with only a few 1000 of photons/pixel/s.
Moreover, this technique possesses an interesting intrinsic scaling advantage when compared to up-conversion imaging approaches. Consider increasing the number of available modes by a factor of $N^2$: both techniques would require factor $N$ increase in the pump laser waist (and thus the crystal to accommodate it). However, to maintain a constant SNR, nonlinear interferometry requires an $N^2$ increase in the pump power, whereas up-conversion imaging would demand an increase of $N^3$.

Whilst the spatial resolution presented here with our basic optical layout is larger than the resolutions anticipated by state-of-the art mid-IR systems (1-10 $\mu$m), the extension to increased imaging capacity is straightforward. The number of available spatial modes is currently limited by the size of the pump waist, and thus the crystal aperture. With an increase in the crystal aperture from $1$ mm to $4$ mm, the number of available spatial modes will grow from $\approx$ 750 to 12,000. Naturally, this 16 fold increase comes at the expense of the per pixel illumination. Small improvements, however, such as optimised camera illumination (2.5 pixels per resolved mode from the existing 5) and increased exposure time (from 1 to 4 seconds) will return us to the current per pixel illumination, while further gains remain accessible via crystal optimisation and increased pump power. These improvements, coupled with more sophisticated approaches to magnification, will yield imaging capacities at the state-of-the-art, while surpassing its speed by leveraging silicon camera technology. Additionally, the latter also ensures the pixel-number of the camera itself is no longer a potential bottleneck, since sensitive Megapixel CMOS-cameras are widely available. 

Crucially for applications in hyper-spectral imaging, the use of a purpose-engineered broadband SPDC source \cite{Vanselow2019Ultra-broadbandPairs} allows a broad wavelength range to probe the sample simultaneously. Harnessing the potential for high spectral selectivity - fundamentally limited only by the bandwidth of the pump laser - will fully unlock the potential of this technique. For example, with efficient extraction of the spectral information and a larger crystal aperture, a mid-IR hyper-spectral microscopy image\footnote{Considering a SNR about 10, 10$\mu$m spatial resolution within a 1 mm FoV (10,000 resolved modes) and 250 spectral modes (a 2.5 $cm^{-1}$ resolution within a bandwidth of 3,4 and 4,3 $\mu$m).} with a 10$\mu$m spatial resolution and 250 spectral modes could be achieved within a few minutes. 

In conclusion, we have experimentally demonstrated that nonlinear interferometry with entangled photons provides a powerful and cost-effective technique for microscopy in the mid-IR, harnessing the maturity of silicon-based near-IR detection technology to allow mid-IR imaging with exceptionally low-light-level illumination. We have shown how this technique can be readily extended to hyper-spectral imaging across nearly 1 $\mu$m. A practical biological sample was imaged with quantum light, revealing morphological features with high resolution. Our results pave the way for broadband, hyper-spectral mid-IR microscopy with fast, wide-field imaging capabilities, enabling far-ranging applications in bio-medical imaging.

\section*{Materials and Methods}

\paragraph*{Experimental Setup} 
A 660 nm continuous wave solid state laser delivers the pump beam for the experiment. Using a telescopic arrangement and a DM, the pump illuminates the ppKTP crystal with a  \( 430\ \mu \)m waist, maximally covering the 1x2 mm crystal aperture. The ppKTP crystal is quasi-phase matched for a collinear type-0 process, specifically engineered to produce highly non-degenerate photon pairs. The photon pairs are ultra broadband due to group velocity matching, with an idler spectrum from 3.38-4.29 $\mu$m. The emerging signal, idler (and pump) fields are subsequently collimated (focused) by an off-axis parabolic mirror (OPM). Using a dichroic mirror (DM), the idler is then split from the signal and pump, and is incident on a sample placed on the end mirror of the interferometer. The end mirrors of the interferometer (and thus the object) are placed in the Fourier plane of the crystal. The use of lenses between the OPM and the idler and signal end mirrors is primarily for obtaining the desired field of view (FoV) on the sample, and for matching the idler and signal interferometric arms, respectively.  In our magnified setup, a 10-fold magnification is introduced in the idler arm via a telescopic arrangement. The pump is reflected using a cold mirror at the focal length of the OPM. The reflected idler, signal and pump fields back-propagate through the same path, back into the crystal, with the pump again generating signal and idler fields. Prior to detection, the signal field is filtered using a 3.5 nm band-pass filter. The Fourier plane of the resulting signal field is then imaged on a CMOS camera. The interferometer arms lengths are required to be matched to within the coherence length of the detected light. For the spectrally filtered image, this length is typically 100 microns. The Michelson-type interferometer geometry strongly simplifies the alignment process. 
\paragraph*{Slide Preparation}
9-12 weeks old C57BL/6J mice were sacrificed by cervical dislocation. For formalin fixation, hearts were removed, rinsed in ice-cold saline and placed in 4\% formalin. After 48 to 72 hrs fixation the tissue was rinsed with (Phosphate-buffered saline) PBS, and finally embedded in paraffin. The paraffin-embedded hearts were cut in transverse sections to a thickness of 2 to 3 $\mu$m and transferred to a low-e slide. Slides were stored at room temperature.



\section*{Acknowledgments}
S.R. acknowledges funding from Deutsche Forschungsgemeinschaft (RA 2842/1-1).

\clearpage

\section*{Supplementary materials}

\subsection*{Working Principle}

Our imaging setup is based on the conceptual arrangement of Michelson-type nonlinear interferometer\cite{Chekhova2016}. A pump beam illuminates a nonlinear crystal twice in sequence, in a folded geometry. Photon pairs are formed in the first and second pass through the crystal, the signal and idler are denoted $\ket{c}_s\ket{d}_i$ and $\ket{e}_s\ket{f}_i$ respectively. After the first path, the idler and signal are split using a dichroic mirror (DM). The idler is sent into a sample with transmittance T and phase shift $\gamma$: $\ket{c}_s\ket{d}_i\to Te^{i\gamma}\ket{c}_s\ket{d}_i+\sqrt{1-T^2}\ket{c}_s\ket{l}_i$, where the state $\ket{l}_i$ encompasses all the loss in the idler arm.
Both idler and signal are then back reflected and aligned to allow minimal distinguishability between the bi-photon amplitudes ($\ket{c}\to\ket{e},\ket{d}\to\ket{f}$) such that after the second pass of the crystal, the obtained state is: 
\begin{equation}
\ket{\phi}= \frac{1}{\sqrt(2)}[(1+Te^{i\gamma})\ket{e}_s\ket{f}_i+\sqrt{1-T^2}\ket{e}_s\ket{l}_i]
\end{equation}
Subsequently, the idler photon is discarded using a DM and the detection probability obtained is: 
\begin{equation}
P=1+T\cos{\gamma}
\end{equation}
Accordingly, an interferometric image with visibility T can be detected. The idler is not detected and the information about the object in the idler path is transferred to the signal through the high spatial correlations shared between the signal and idler modes. This feature when combined with non-degenerate downconversion allows sensing and detection at different wavelength ranges.

\subsection*{Theoretical Model}

To ascertain the theoretical imaging capacity of our implementation we developed a simple theoretical framework to calculate both the anticipated field of view (FoV) and resolution of our optical system. We also make a comparison to the theory of the SPDC wavefunction. 

The FoV in our imaging system is impacted by the emission angle of the down-converted idler light and any subsequent magnification that defines the size of the illuminating spot
\begin{equation}\label{eq.3}
{\rm FoV}=\frac{2f\tan(\theta_i)}{M}\approx\frac{2f\theta_i}{M},
\end{equation}
where $f$ denotes the focal length of the collimating optical element adjacent to the crystal, $\theta_i$ is the divergence angle \footnote{All widths are specified as half-width half-maximum (FWHM).}, and $M$ represents the magnification of the optical system after the collimation. The opening angle, $\theta_i$, results from the phase-matching conditions
\begin{equation}\label{eq.4}
    \theta_i=\lambda_i\sqrt{\frac{2.78}{L\pi}\frac{n_in_s}{\lambda_in_s+\lambda_sn_i}},
\end{equation}
where $L$ denotes the crystal length, $\lambda_i$ and $\lambda_s$ are the wavelengths of the idler and signal respectively and $n_i$ and $n_s$ are the reflective indices of KTP at the designated wavelengths. 

The resolution is limited by two constraints. The first limitation arises from the standard diffraction limit, applicable to most conventional imaging and microscopy techniques,
\begin{equation}
\delta x_{NA}=\frac{\lambda}{2NA},
\end{equation}
where $\lambda$ and $NA$ are the wavelength and numerical aperture of the optical elements in our system.

The second condition upon the resolution - and here the limiting condition - is an expression of the strength of momentum (or position) entanglement inherent to the bi-photon state \cite{Moreau2018,BrambillaSimultaneousDown-conversion} and is given by 
\begin{equation}\label{eq6}
\delta x_{corr}=\frac{\sqrt{2\ln{2}}f\lambda_i}{\pi w_pM},
\end{equation}
where $w_p$ denotes the pump waist. 
The number of spatial modes is therefore 
\begin{equation}\label{eq7}
N_{modes}=\left({\frac{{\rm FoV}}{\delta x_{corr}}}\right) ^2=\frac{5.56\pi}{\ln2}\frac{{w_p}^2}{L}\frac{n_in_s}{\lambda_in_s+\lambda_sn_i}
\end{equation}
Unsurprisingly, the number of spatial modes does not depend on the magnification ($M$) - provided one has a sufficiently large NA.  By substituting the experimental values into Eq.\ref{eq7} we obtain $911 \pm 13$ spatial modes.

\subsection*{Derivation of the emission angle} 
The width (FWHM) of the idler in the far field corresponds to an emission angle, ${\rm sinc}^2(\frac{\Delta kL}{2})=\frac{1}{2}$, corresponding to a phase mismatch $\Delta k L=2.78$. Expressing the phase mismatch in terms of the transverse emission angle we obtain
\begin{equation}\label{phase_match}
(\frac{\pi n_i}{\lambda_i}\tilde{\theta_i}^2+\frac{\pi n_s}{\lambda_s}\tilde{\theta_s}^2)\frac{L}{2}=1.39.
\end{equation}
Assuming transverse momentum conservation, the FWHM of the idler emission angle within the crystal is given by
\begin{equation}
\tilde{\theta_i}=\sqrt{\frac{2.78}{\pi L}\frac{n_s{\lambda_i}^2}{n_s n_i\lambda_i+{n_i}^2\lambda_s}},
\end{equation}
corresponding to angle in free space of
\begin{equation}
\theta_i=n_i\cdot\tilde{\theta_i}=\lambda_i\sqrt{\frac{2.78}{\pi L}\frac{n_s n_i}{n_s \lambda_i+n_i\lambda_s}}.
\end{equation}
\subsection*{Experimental values}
The signal wavelength was measured using a spectrometer to be $\lambda_s = 801 \pm 1 nm$. The pump wavelength was measured using a wave meter and the value obtained is: $\lambda_p = 659.75 \pm 0.01 nm$. The idler wavelength was calculated according to energy conservation ($\lambda_i=\frac{1}{\frac{1}{\lambda_s}+\frac{1}{\lambda_p}}$), the obtained value is: $\lambda_i = 3.74 \pm 0.02 \mu$m. The refractive indexes were calculated using the appropriate Sellmeier equations\cite{Katz2001,Fan1987} for KTP: $n_s= 1.845, n_i= 1.752$. The length of crystal is: $L= 2$mm and the pump waist at the crystal is $w_p=431 \pm 6 \mu$m. 

The SNR was calculated for the difference image, by dividing the number of mean counts by the standard deviation within a 7x7 pixel region. The size of the region was chosen to be comparatively small in order to exclude the effect of the varying illumination profile. 

The power on the sample was measured by detecting the signal light on the camera after the first path of the pump through the crystal. The number of counts on the camera was converted to the the number of electrons according to a constant conversion factor (provided by PCO) and then to the number of photons using the quantum efficiency of the CMOS detector at 800 nm ($N_{photons}=\frac{0.46N_{counts}}{0.42}$). The accordingly obtained number of $1.5\cdot10^8$ photons at a wavelength of 3.5 $ \mu m$ corresponds to sample illumination power of 17 pW.
Given the bandwidth of the photons of around $1.5\cdot10^13$ Hz, the brightness could be increased by more than 4 orders of magnitude without leaving the low-gain regime of SPDC. In the high-gain regime on one hand spectral and spatial mode numbers would start to decrease, while on the other hand quantum-enhanced phase-sensitivity effects would become relevant.  

\subsection*{Future setup optimization}
A future increase in the crystal size from 1 to 4 cm will allow the increase of total spatial modes by $4^2=16$, according to Eq. 7. To attain a SNR of 10, while maintaining the same pump power, we plan to optimize the number of pixels per resolvable elements and reduce it by 2 per direction (from 5 down to 2.5), corresponding to a total factor of 4. Also, by improving the visibility for the magnified setup and reaching the unmagnified visibility corresponding to a SNR of 25, we will be able to reduce the current exposure time from 1 to 0.65 seconds ($1 sec\cdot \frac{16}{(25/10)^2\cdot4}$). For a spectral resolution of $2.5 \:cm^{-1}$ and a spectral range of $625 \:cm^{-1}$, a hyperspectral image with 250 spectral modes can be recorded in just 2.7 minutes. 

\subsection*{Theoretical formulation from the bi-photon state}
In the low gain limit, the bi-photon state produced by the SPDC can be written in the angular spectra representation
\begin{align}
|\psi_1 \rangle = \int C_1({\bf q}_i,{\bf q}_s) |{\bf q}_s, {\bf q}_i \rangle ,
\end{align}
where  ${\bf q}_s$ and ${\bf q}_i$ represent the transverse components of the SPDC emission.
Analogous to the spectral properties of SPDC emission, the spatial properties are governed by the intersection of energy and momentum conservation,
\begin{align}
C_1({\bf q}_i,{\bf q}_s) = \alpha({\bf q}_s,{\bf q}_i) \phi( {\bf q}_s, {\bf q}_i), 
\end{align}
specified by the $ \alpha({\bf q}_s,{\bf q}_i)$, the pump function and $\phi( {\bf q}_s, {\bf q_i})$ is the phase-matching function, respectively. 
We recast longitudinal phase mismatch $\Delta k_{z_0}$ in terms of the transverse momenta
\begin{align}\Delta k_z = \Delta k_{z_0}-\tfrac{1}{2}(\tfrac{|{\bf q_s + q_i }|^2}{ k_p} + \tfrac{|{\bf q_s }|^2}{ k_s} + \tfrac{|{\bf q_i }|^2}{ k_i})\end{align}
where $\Delta k_{z_0}$ is the residual longitudinal collinear phase-mismatch at the given emission wavelength. The phase function is given by 
\begin{align} \phi( {\bf q}_s, {\bf q}_i) = \mathrm{Sinc}\left( \frac{L \Delta k_z}{2} \right), \end{align}
and the pump function is 
\begin{align} \alpha ( {\bf q}_s, {\bf q}_i)= \exp{(-| {\bf q}_s+{\bf q}_i |^2 \omega _p^2 /4)}, \end{align}
where $\omega _p$ is the pump waist. Consider two identical SPDC processes aligned in series with all modes matched
\begin{align} | \Psi \rangle =& \tfrac{1}{\sqrt{2}}(|\psi _1 \rangle + \psi _2\rangle) \end{align}
More explicitly,
\begin{align} |\Psi \rangle =\tfrac{1}{\sqrt{2}} \int d{\bf q}_s d{\bf q}_i C_1({\bf q}_i,{\bf q}_s)|{\bf q}_i,{\bf q}_s\rangle  + \tfrac{1}{\sqrt{2}} \int d{\bf q}_s d{\bf q}_i \exp{i\phi} \,C_1({\bf q}_i,{\bf q}_s)|{\bf q}_i,{\bf q}_s\rangle \end{align}
If we want to consider an overall transmission, $\eta$, of the idler between the first and second crystal, we considering an additional loss mode $\mathbf{q}_{i}^{\star}$. Considering an intensity measurement in the Fourier plane, one obtains
\begin{align} \langle \Psi|{\bf q}^{\dagger}_s {\bf q}_s |\Psi \rangle = \frac{1}{2}(\eta + 2 \sqrt{\eta} \cos{\phi} + 1 )\int d{\bf q}_i d {\bf q}_s  |C_1({\bf q}_i ,{\bf q}_s )|^2  \end{align} 
If we assume no longitudinal phase mismatch at all wavelengths, then the number of modes decreases linearly with increasing wavelength (where one has chosen to fix the pump wavelength). This is simply a consequence of increasing non-degeneracy reducing the spatial entanglement of the system. 
For the unmagnified configuration and our earlier specified system values, the conditional probability density gives a resolution of $326\mu$m (FWHM). The theoretical FoV is 9.0 mm (FWHM). The Schmidt decomposition of the joint probability density gives an effective mode number of 500. The discrepancy between the Schmidt number and the mode number obtained earlier, reflects that the latter considers the ratio of a Sinc width to a Gaussian width. 

\clearpage

\bibliographystyle{Science}

\end{document}